\documentclass[runningheads]{llncs}
\usepackage{nicefrac}
\usepackage{amssymb}
\usepackage{booktabs}
\usepackage{graphicx}
\usepackage{color}
\begin{document}
\title{Virtual Thin Slice: 3D Conditional GAN-based Super-resolution for CT Slice Interval}
\titlerunning{Virtual Thin Slice}
\author{Akira Kudo\inst{1} \and
Yoshiro Kitamura\inst{1} \and
Yuanzhong Li\inst{1} \and
Satoshi Iizuka\inst{2} \and
Edgar Simo-Serra\inst{3}}

%\authorrunning{Anonymous et al.}
\authorrunning{A. Kudo et al.}
\institute{Imaging technology center, Fujifilm corporation, Minato, Tokyo, Japan \and
Center for Artificial Intelligence Research, University of Tsukuba, Tsukuba, Ibaraki, Japan \and
Department of computer science and engineering, Waseda university, Shinjuku, Tokyo, Japan \\ 
\email{akira.kudo@fujifilm.com}
}
\maketitle              % typeset the header of the contribution
\begin{abstract}
Many CT slice images are stored with large slice intervals to reduce storage
size in clinical practice. This leads to low resolution perpendicular
to the slice images (\textit{i.e.}, z-axis), which is insufficient for 3D visualization
or image analysis. In this paper, we present a novel architecture based on
conditional Generative Adversarial Networks (cGANs) with the goal of
generating high resolution images of main body parts including head, chest, abdomen and legs. However, GANs are known
to have a difficulty with generating a diversity of patterns due to a phenomena
known as mode collapse. To overcome the lack of generated pattern
variety, we propose to condition the discriminator on the different
body parts. Furthermore, our generator networks are extended to be
three dimensional fully convolutional neural networks, allowing for the
generation of high resolution images from arbitrary fields of view.  In our
verification tests, we show that the proposed method obtains the best
scores by PSNR/SSIM metrics and Visual Turing Test, allowing for accurate reproduction of
the principle anatomy in high resolution.  We expect that the proposed method
contribute to effective utilization of the existing vast amounts of thick CT
images stored in hospitals.

\end{abstract}

\section{Introduction}
\begin{figure}[t]
\centering
\includegraphics[width=\linewidth]{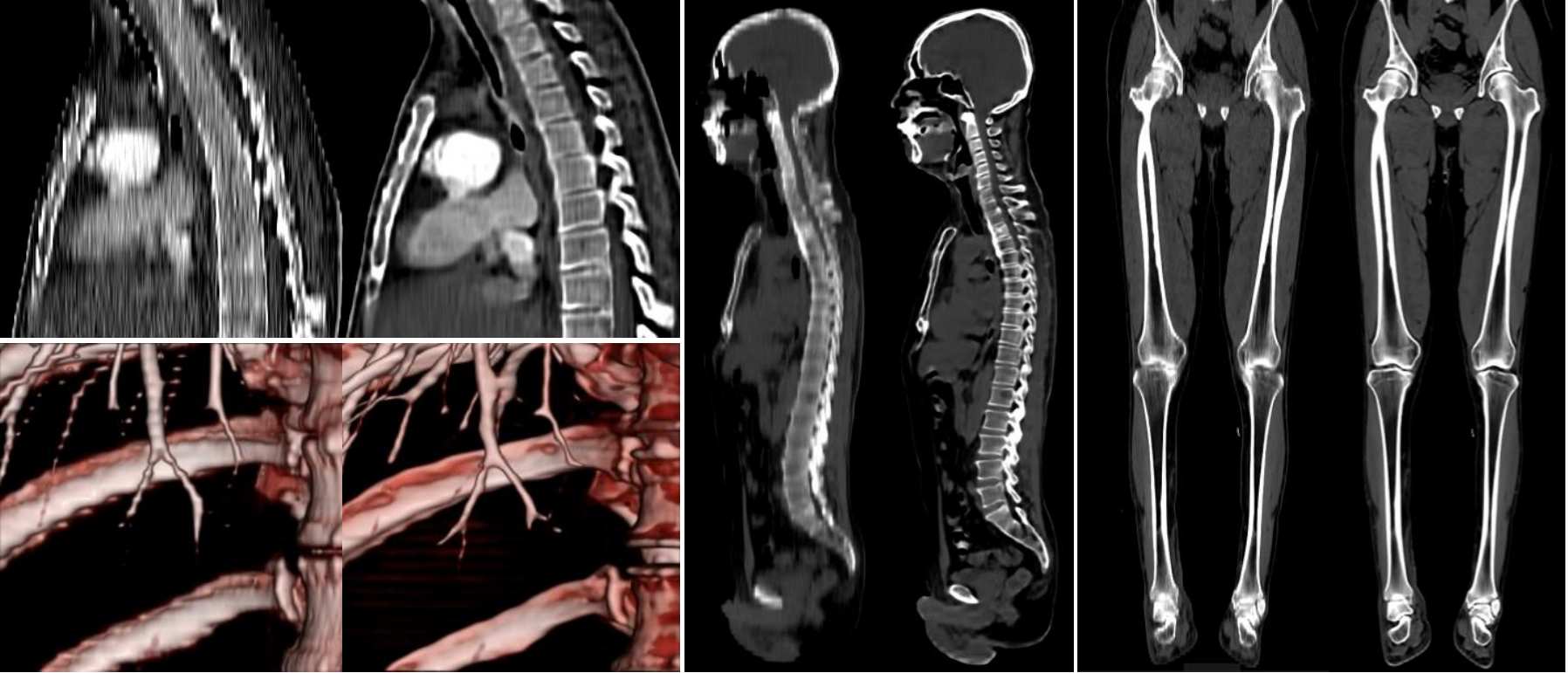}
\vspace{-8mm}
\caption{Comparison of original thick image and the virtual thin image output generated by the proposed approach.
On the left top row, the CT sagittal view of original thick image is blurred
with each vertebrae bone being nearly indistinguishable, while
they become clear in corresponding $\times$8 super resolution image (Virtual
Thin). Arbitrary size data, even the whole body shown on the right, is available for inputs and 
capable of reconstructing natural image regardless of body part.
On the left bottom row, fine blood vessel are smoothly reconstructed in a
volume rendering view.} \label{fig:intro}
\end{figure}

Image diagnosis plays an important role in recent healthcare solutions. The
quality of diagnostic images largely affects the quality of diagnosis. The
images such as CT or MRI acquired in hospitals are normally stored in Picture
Archiving and Communication Systems (PACS). Although thin slice images,
with slice intervals are about less than 1 mm, are frequently used for
diagnosis, thick slice images with large slice intervals are
used for long term storage to reduce the data size. However,
the stored thick slice images do not have sufficient resolution
for sagittal or coronal views, and also have limited applicability to 3D
visualization (volume rendering). To address this, we present a novel super
resolution algorithm for CT images based on Generative Adversarial
Networks (GAN). Our goal is to generate high resolution 3D images
corresponding from the input thick slice images. We base our
approach on adversarial training~\cite{goodfellow2014generative}
and aim to generate realistic-looking high-resolution CT images.
One of the major difficulties is that CT images can be very
diverse (\textit{e.g.}, imaged body part, voxel size, resolution, slice thickness,
slice interval, etc…), which can be difficult to synthesize with
GANs. This difficulty is due to a phenomena known as mode collapse, in which
the model becomes only able to synthesize a small subset of the original
training data and presents a significant decrease in the output diversity~\cite{metz2016unrolled}.
We overcome this issue by additional conditioning of the discriminator
on additional information, and use a three dimensional fully convolutional
network to synthesize the high resolution CT images.
Figure~\ref{fig:intro} shows example input thick images and
the corresponding thin images synthesized by the proposed Virtual Thin Slice (VTS) method. The
vertebrae bone structure is clearly reconstructed on the sagittal view, and
fine blood vessels are reproduced well on the VR image.

\section{Related work}
Single image super resolution is a major problems in computer vision field
with a long history.
The very first approach were filtering approaches, such as linear, bicubic or Lanczos filtering~\cite{duchon1979lanczos} which do not require huge computation. 
Yang et al.~\cite{yang2014single} categorized super resolution technique into 4
groups, which are prediction models, edge-based methods, image statistical
methods and path-based methods. Recently, deep convolutional neural networks
(CNN) based methods are showing significant performance in image recognition
area~\cite{radford2015unsupervised}.  SRCNN~\cite{dong2014learning} improved
performance on 2D image super resolution tasks by training non-linear
low-resolution to high-resolution mappings using CNN filters. This is achieved
through the minimization of pixel-wise Mean Squared Error (MSE) between
reconstructed image and ground truth high resolution image. However, pixel-wise
losses cannot capture perceptual differences~\cite{blau2018perception}, thus
the output tends to be blurred and look unrealistic to human eyes.
Adversarial training schemes~\cite{goodfellow2014generative} give much sharper
result in image conversion task. 
Our work builds upon adversarial training to
obtain better results. 
Some research focus on stabilize GAN training to reduce mode collapse, while  
there is a drawback of the computational cost~\cite{metz2016unrolled}.

Conditional GAN~\cite{mirza2014conditional} is a conditioned min-max game
between generator and discriminator. Isola~\cite{isola2017image} proposed
Pix2Pix algorithm using images as conditional information, using pair of input
image and target ground truth image. 
While adversarial approach gives more
adequate to human perception, there are trade-offs between the perception and
distortion of generating images~\cite{ledig2017photo}. 

In medical imaging, few researchers work on the 3D image super resolution. Chen
et al. targeted 3D MRI super-resolution for medical image
analysis~\cite{chen2018efficient}. Their target was limited to brain images.
One of our contributions is realizing 3D CT image super resolution for any kind
of body parts with a single generator network. Another contribution is the
conditioning of the discriminator on the different body parts inspired by
conditional GAN, and the ability to perform super-resolution of  3D medical
images of arbitrary sizes.

\section{Method}

\subsection{Objective Function}

\begin{figure}[h]
\centering
\includegraphics[width=\linewidth]{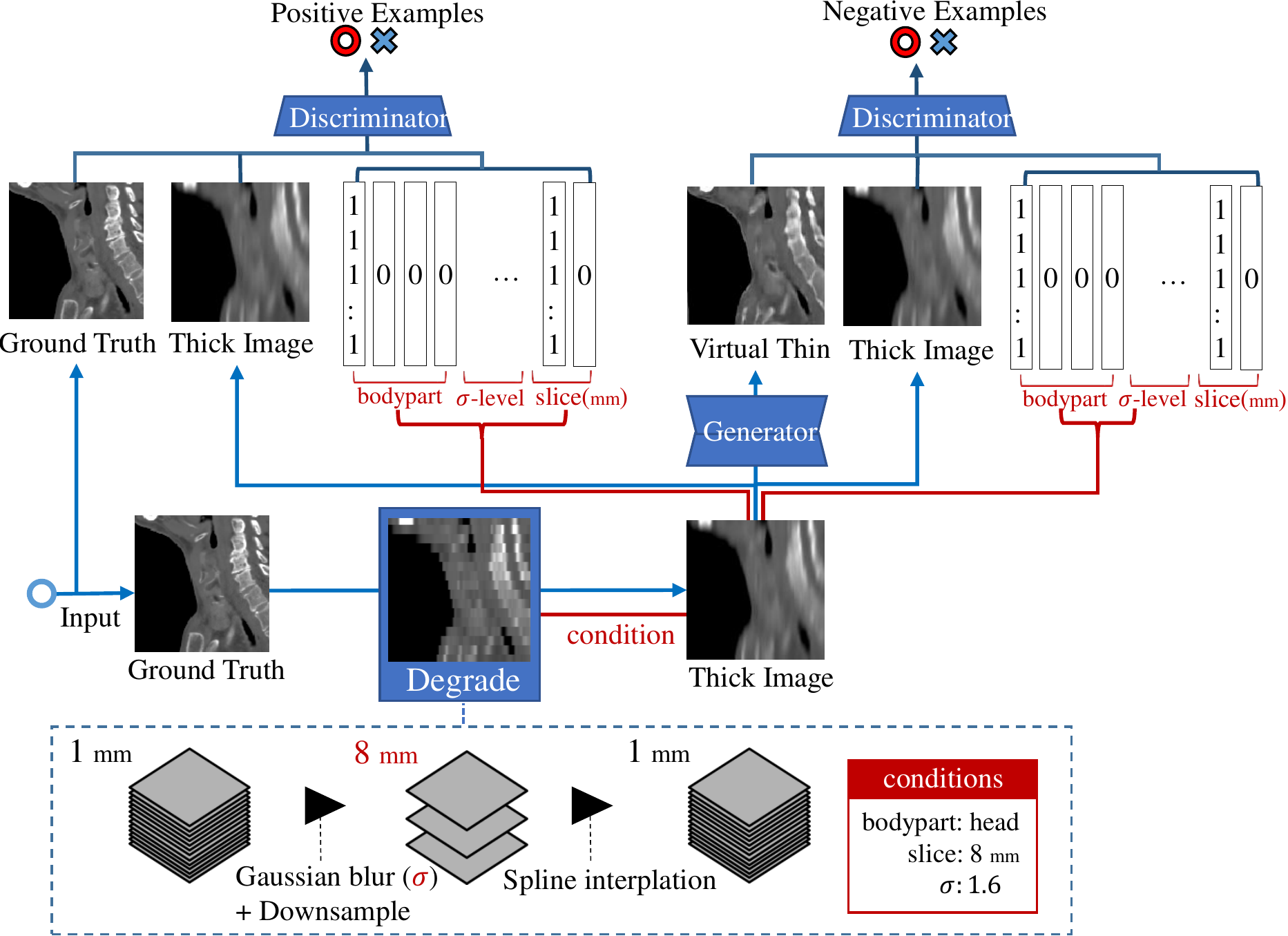}
\vspace{-2mm}
\caption{Adversarial training framework of thick-thin slice translation on CT images. In each training iteration,
 the thin slice input data is randomly degraded to simulate thick slice data. 
For the Generator input, we feed 1 mm spline interpolated 3D thick slice image itself. 
On the other hand, for the Discriminator input, we feed generated Virtual Thin image with slice condition including body part information with degraded parameter scales.  
} \label{fig:frameworks}
\end{figure}

Our approach is based on conditional GAN, using pairs of low resolution data
and high resolution data with slice information.  The objective is to learn the
transformation from of the thick slice image $x$ to the virtual thin slice
image $y$.
Additionally, the discriminator is condition on a vector $w$,
allowing the objective function to be expressed as
\begin{equation}
\label{eq:a}
	L_{cGAN}(G,D) = E_{(x,y)}\left[\,\log D(x,y,w)\,\right]+E_{x}\left[\,\log (1-D(x,G(x),w))\,\right].
\end{equation}
\noindent where the model $G$ tries to minimize this objective against an adversarial model $D$
that tries to maximize it. Both $G$ and $D$ can be implemented as Convolutional
Neural Networks (CNN).
We also use a $L_1$ loss to calculate pixel-wise appearance differences between
ground truth images and generated images, which has been shown to give less blurring than the $L_2$ loss in a diversity of image-to-image translation tasks~\cite{isola2017image}.
Therefore, our final objective is
expressed as
\begin{equation}
\label{eq:b}
	G^* = \arg \min_{G} \max_{D} \; L_{cGAN}(G,D) + \lambda L_{L1}(G) 
\end{equation}
Figure~\ref{fig:frameworks} illustrates the proposed adversarial training
procedure. Note that the additional conditions are not inputted into the
generator.

\subsection{Network Architecture}
\begin{figure}[h]
\centering
\includegraphics[width=\linewidth]{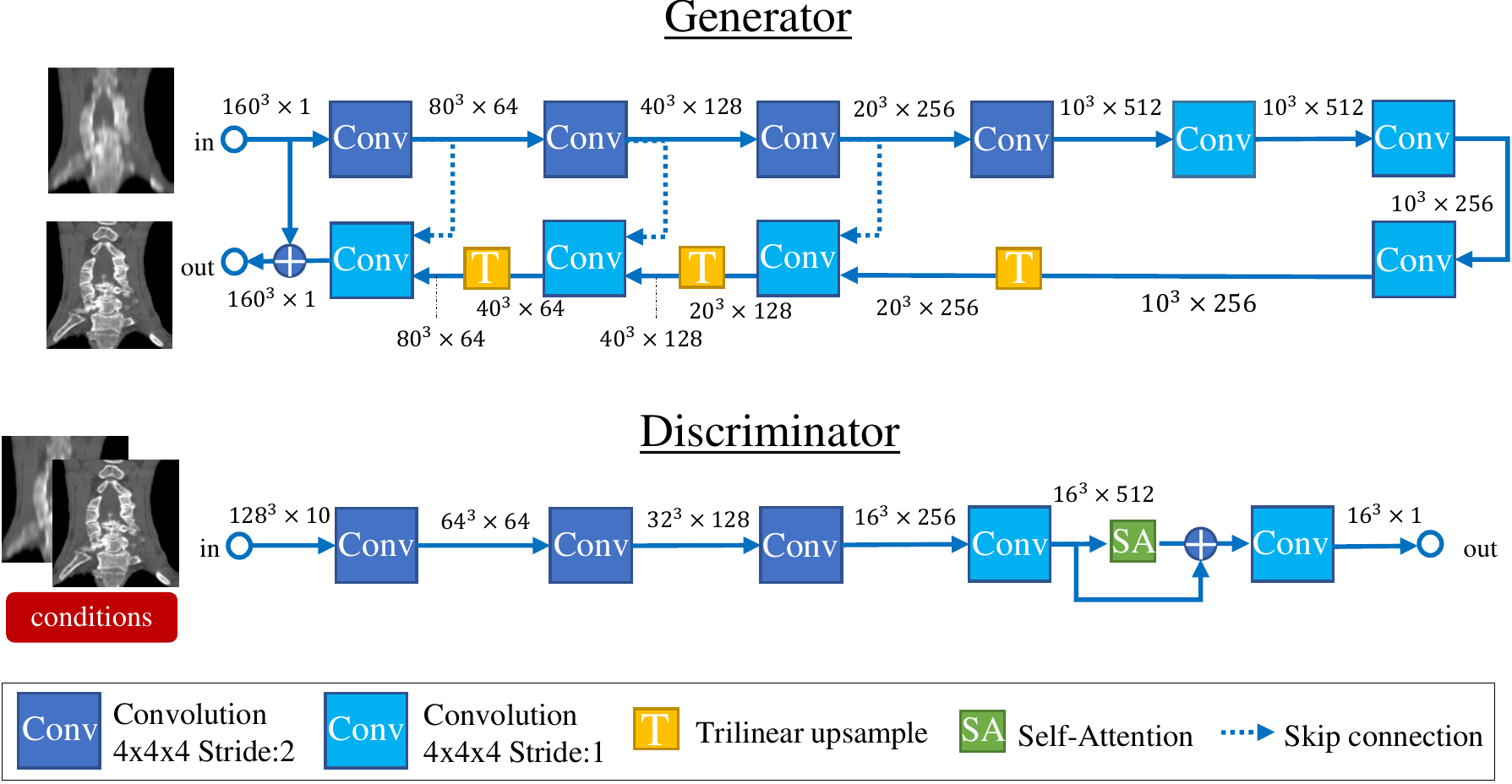}
\vspace{-2mm}

\caption{Overview of the generator and discriminator network architectures. ($x^3\times c$)
denotes the size of 3D feature map volumes, where $c$ denotes the number of
channels. } \label{fig:architecture}
\end{figure}

Both our generator and discriminator models are based on Convolutional Neural
Networks.
Each convolutional layer consists of $4\times4\times4$
sized kernel followed by batch normalization~\cite{ioffe2015batch} and
LeakyReLU ($\alpha=0.2$) as the activation function.
Instead of max-pooling,
strided convolution are used and image resolution is reduced to
$\nicefrac{1}{2}$ in each encoder convolutional layer.
We set 64 channels for the first layer for both of the generator and the
discriminator to get sufficient quality results and acceptable computation
time.
Figure~\ref{fig:architecture} illustrates the architecture of our generator and
discriminator networks.

\subsubsection{Generator}
The generator uses an encoder-decoder type architecture inspired by
U-Net~\cite{ronneberger2015u}. 
The resolution is decreased 4 times such that the minimum feature map size is
$\nicefrac{1}{16}$, and restored to the original size with trilinear interpolation.
We used trilinear interpolation instead of transposed convolution for up-sampling to avoid generating checkered patterns due to uneven overlap of the kernel. 
In each convolutional layer, more feature channels generate better results,
but require more resources and computational time.
The generator estimates the high frequency components and the output is finally added to the input image. 

\subsubsection{Discriminator}
For the discriminator input, thick image (1 channel), thin image (1 channel)
and slice information (8 channels) is given. 
Also a self-attention layer~\cite{zhang2018self} is added in the fourth layer of the network.
Self-attention mechanism is an idea to introduce global information between
layers by computing attention maps which show the relevant area. 
In our case, self-attention did not largely affect the final performance, however, it speeded up the convergence of adversarial training.
Final output is converted to a probability with a Sigmoid function.

\subsection{Training data}
We introduce a Degrader procedure to randomly degrade the original thin image
to thick slice image. In the Degrader, the input 3D image is down-sampled with
Gaussian smoothing and spline interpolation is used to generate the missing
slices. To simulate various combination of slice thickness and slice interval,
the number of slices is reduced to either $\nicefrac{1}{4}$ or
$\nicefrac{1}{8}$, then spline interpolation and random Gaussian noise is
applied. The training samples are randomly cropped from the training images
with an affine transform.

\subsection{Conditioning Vector}
We condition the discriminator on a vector $w$, containing various information
about the input image. In particular, the type of input data (head, chest,
abdomen or leg) is provided, in addition the slice interval (4mm or 8mm),
and the scale of the standard deviation $\sigma$ (2 scales) used for the Gaussian kernel
which is treated as the slice thickness. In total, 8 channels are added as
conditional information to the discriminator.

\section{Experiments}
For verification of the proposed method, we use Peak Signal-to-Noise Ratio
(PSNR) and Structural Similarity Index Metric (SSIM)~\cite{wang2004image} as
automatic metrics.  There is a huge gap between PSNR and human perception
sense, thus do not always present reasonable result. SSIM is more reasonable
measure in super resolution field, however there is still a gap to human
perception sense.  Therefore, as additional experiment, we conducted Visual
Turing Test (VTT). We asked 8 people who is either radiology technician or
medical image research scientist to select the most high visibility image
generated from 4 different methods each input. In the VTT, 4 images are shown
in random order for 50 times to aggregate the answered ratio.

\subsection{Datasets and  Data Augmentation}
We prepared 354 CT data (head:99, chest:98, abdomen:100, legs:57) for training which are obtained from diverse manufacturer's equipments (\textit{e.g.}, GE, Siemens, Toshiba, etc…). 
They have been carefully selected to not contain metal artifacts or noises
because the discriminator is prone to reproduce such artifacts. 
The input images CT values are clipped to the [-2048, 2048] range and then normalized to be in the [-1, 1] interval.
In general, thick CT images are acquired in the range of 3-10 mm interval. On the other
hand, 1mm slice interval is enough for 3D visualization of principle anatomy by
volume rendering. We set the experimental setting to generate 1 mm slice
interval images from 8 mm. Therefore our datasets are only data with smaller
than 1.0 mm slice interval. All images are rescaled to 1mm isotropic voxels in
preprocessing steps.
In each training iteration, we randomly crop $160\times160\times160$ voxels from the input data and apply data augmentation.
In particular, we apply an affine transformation consisting of a random rotation between -5 and +5 degrees, and random scaling between -5\% and +5\%, both sampled from uniform distributions.
The generated thick slice image inputs are subject to Gaussian filtering with $\sigma$ uniformly sampled between 0.0 and 3.2 voxels before being downsampled to 1/4 or 1/8 resolution. 
As test datasets, we prepared 53 CT data (head:12,
chest:16, abdomen:15, leg:10). 

\subsection{Results}
We report for reference methods adapted to 3D, including bicubic,
SRCNN~\cite{dong2014learning}, Pix2Pix~\cite{isola2017image}, and Virtual Thin
Slice (VTS, our approach). For our approach we perform an ablative study where
we remove the conditional vector from the discriminator and high-frequency
component prediction.  We employ SRCNN's 3 layer 9-1-5 model with each kernel
expanded to 3D. Pix2Pix's convolutional networks are also replaced to
$4\times4\times4$ sized kernels in each layer to adapt 3D and iterates down
sampling until the feature image size become one pixel. Each type of network
architecture is trained for around 100 epochs with Adam optimizer having
learning rate for $2\times10^{-4}$ and momentum parameter $\beta_1=0.5$. 

Table~\ref{tab1} shows average PSNR and SSIM calculated over the test datasets.
VTS has the highest score among other methods and the ablation study shows that
removing either the conditional vector or high-frequency prediction lowers the
quality of the generated outputs.  

\begin{figure}[h]
  \centering
    \begin{tabular}{c}
       \begin{minipage}{0.725\hsize}
          \includegraphics[width=1.0\linewidth]
                          {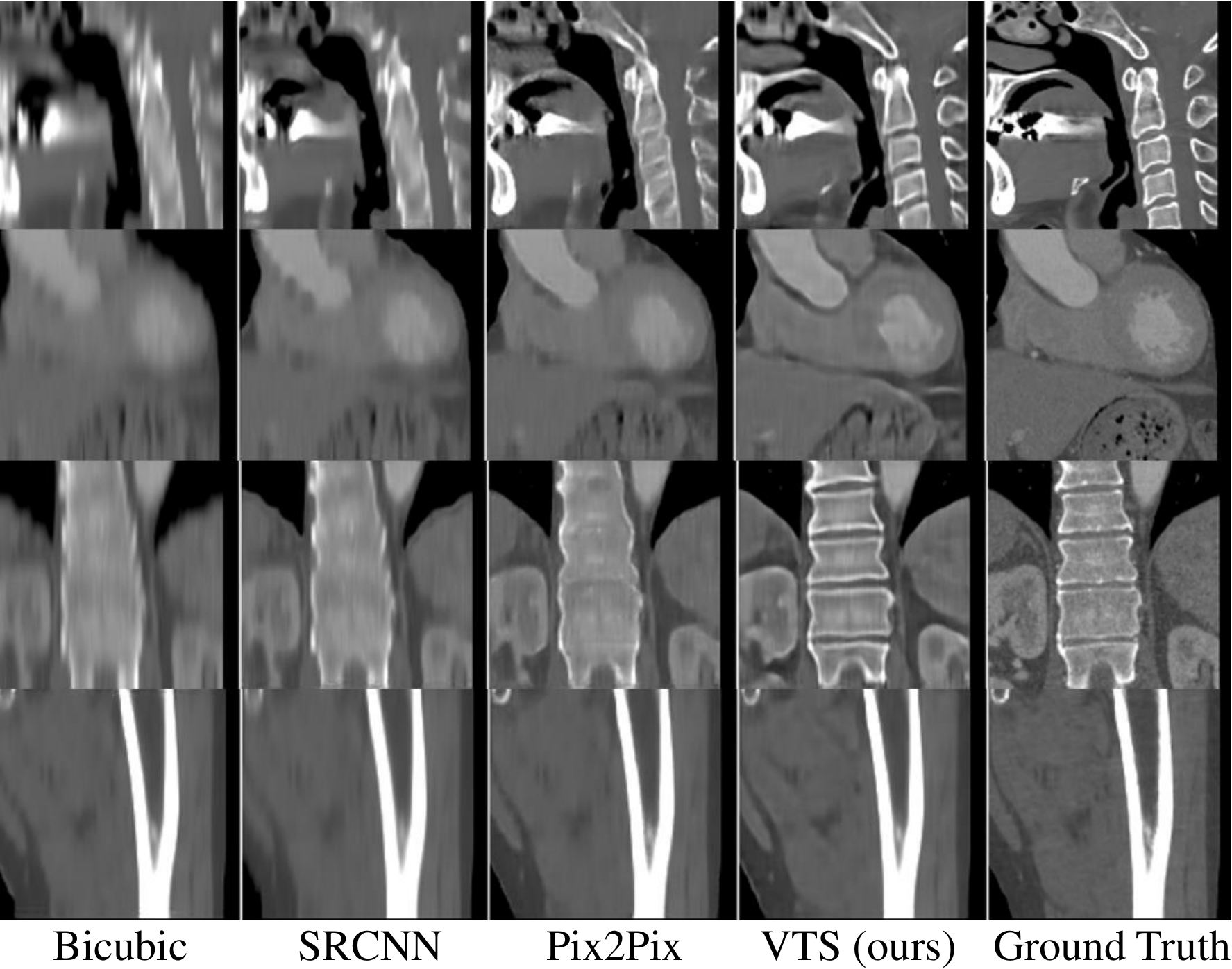}
       \end{minipage}
      \begin{minipage}{0.265\hsize}
        \centering
          \includegraphics[ width=1.0\linewidth]
                          {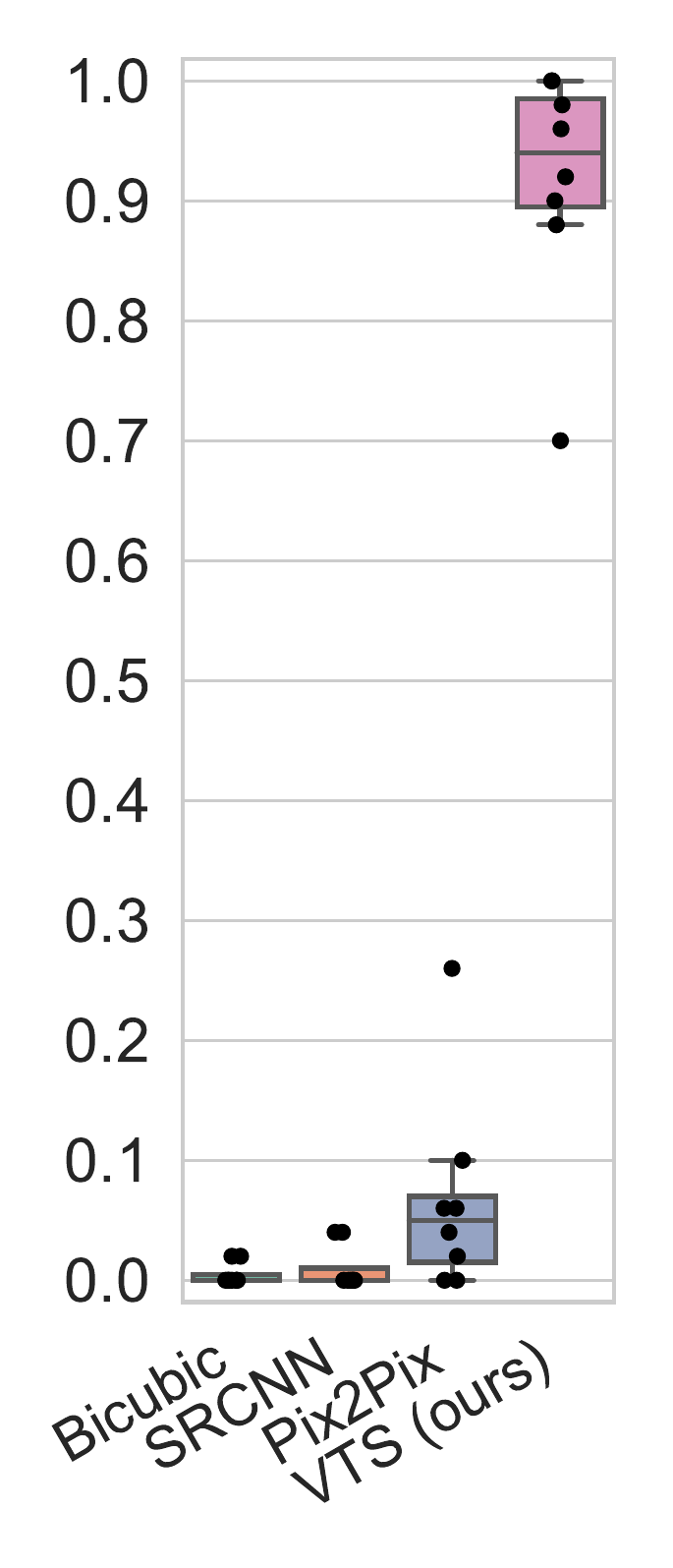}
      \end{minipage} \\ [-7pt]
 \begin{minipage}{.71\hsize}
        \caption{Comparison of the generated images of bicubic,
SRCNN~\cite{dong2014learning}, pix2pix~\cite{isola2017image}, VTS (ours) and
corresponding ground truth thin slice image. [ $8\times $ slice interpolation]}
\label{fig:results}
    \end{minipage}%
    \hfill%
	 \begin{minipage}{.25\hsize}
		\vspace{0mm}
        \caption{The answered ratio in the Visual Turing Test.}
\label{fig:vtt_results}
    \end{minipage}%
    \end{tabular}
\end{figure}

\begin{table}[t]
\caption{PSNR and SSIM comparison result in experiments.}\label{tab1}
\centering
\setlength{\tabcolsep}{6pt}
\vspace{-2mm}
\begin{tabular}{rcccccc}
\toprule
Methods & GANs & Conditional? & HF prediction? &  PSNR & SSIM \\
\midrule
Bicubic & &  & & 32.34 & 0.878 \\
SRCNN~\cite{dong2014learning} & &  & & 33.73 & 0.904 \\
Pix2Pix~\cite{isola2017image} & \checkmark& & & 35.14 & 0.925 \\
\midrule
\bf{VTS (ours)} & \bf{\checkmark} & \bf{\checkmark} & \bf{\checkmark} & \bf{35.73} & \bf{0.933} \\
(w/o) condition & \checkmark&& \checkmark & 35.17 & 0.924 \\
(w/o) HF pred. & \checkmark&\checkmark & & 33.70 & 0.905 \\
\midrule
Ground Truth & -& -& -& $\infty$ & 1.000 \\
\bottomrule
\end{tabular}
\end{table}
As we can see in Figure~\ref{fig:results}, proposed VTS model generated the
best perceptual quality with more sharpness and realistic images than other
models. 
In particular, VTS works better with high intensity values such as bone boundary area rather than soft tissues.
Although Pix2Pix model has similar PSNR/SSIM score to VTS, VTS was preferred roughly 90\% of the time in VTT presented as shown in Figure~\ref{fig:vtt_results}. 
The boxplot shows the answered ratio among 4 methods by the  research participants.
The images in which Pix2Pix was preferred over VTT consists primarily of legs data which have small difference between thick and thin as shown bottom row in Figure~\ref{fig:results}.
Even with some test data containing metal artifacts and unknown test
patterns, the generated images are consistent with the input patterns, and don't contain
enhancing artifacts or noise.

Another important feature of the proposed method
is that the generator network is a fully convolutional neural network, and as such can handle each part of
body and also field of view. 
Additionally, we have performed a verification test on 66 real thick slice images covering a wide condition of view with varying slice numbers (10 to 327), slice intervals (from 3.0 to 10.0 mm), and FOV (128 to 512 mm*mm). Using this dataset, we confirmed that the generator networks are able to successfully generate 1mm slice interval images from the diversity of slice spacing and FOV images. We attribute these results to the wide range of data augmentation that we apply during training. 
Example of generated HR image from real existing thick slice image for either whole body or chest are shown in Fig.~\ref{fig:intro}. The entire images are naturally
reconstructed with no seams.

\section{Conclusion}
In this paper, we have presented a super resolution algorithm that can be applicable
for CT images of main body parts and various field of view. By inputting additional
information regarding input data in the discriminator network, we show that
output data quality increases significantly. Furthermore, the additional
information is not necessary as test time.
Numbering vertebrae bone is clearly easier with our VTS images compared to the
original thick images. 
Also, we believe in-depth evaluation on abnormal images is an important next step for future work.
In the future, we expect our VTS method will take on a
role for the further development of medical image analysis and diagnosis
support tasks, such as bone labeling and lung section segmentation, for thick
slice data.

%
% the environments 'definition', 'lemma', 'proposition', 'corollary',
% 'remark', and 'example' are defined in the LLNCS documentclass as well.
%

\bibliographystyle{splncs04}
\bibliography{mybibliography}

\section*{Acknowledgements}

We acknowledge using the Reedbush-L (SGI Rackable C2112-4GP3/C1102-GP8) HPC system in
the Information Technology Center, The University of Tokyo for the GPU computation required in this work.

\newpage
\section*{Appendix}
\begin{figure}[h]
\centering
\includegraphics[width=0.91\linewidth]{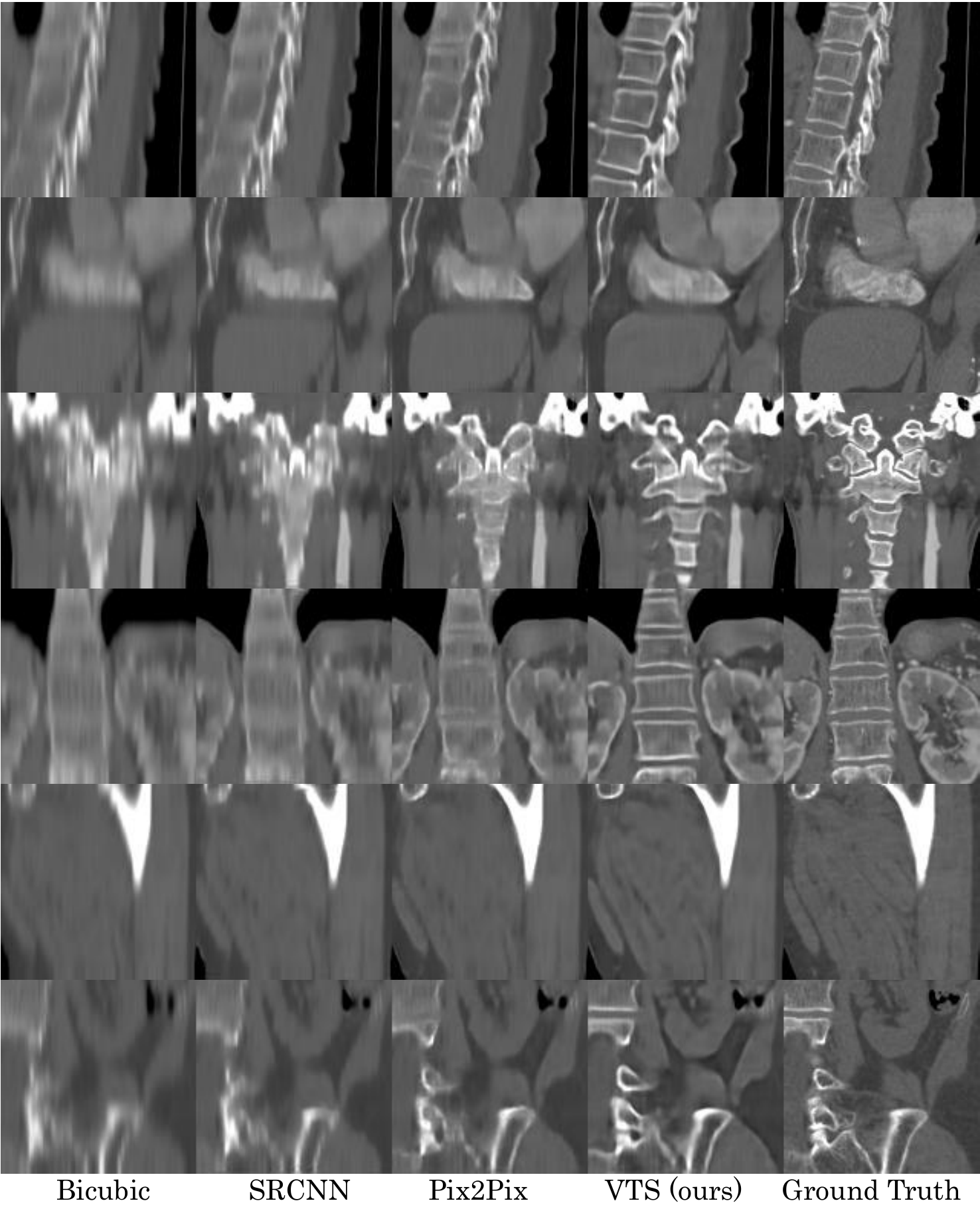}
\vspace{-3mm}
\caption{Results using bicubic,
SRCNN~\cite{dong2014learning}, pix2pix~\cite{isola2017image}, VTS (ours) and
corresponding ground truth thin slice image. } \label{fig:comparison}
\end{figure}

We include a variety of additional generated images from proposed VTS and other methods in Figure~\ref{fig:comparison}.

\end{document}